\documentclass[11pt]{article}
\usepackage[a4paper,left=28mm,right=28mm,top=25mm,bottom=25mm]{geometry}
\usepackage[T1]{fontenc}
\usepackage[utf8]{inputenc}
\usepackage{microtype}
\usepackage{graphicx}
\usepackage{amsmath,amssymb,amsthm,amsfonts,mathtools}
\usepackage{bm,booktabs,array,float,needspace,longtable}
\usepackage{mathrsfs}
\usepackage{cite}
\usepackage{xcolor}
\usepackage{xcolor}

\definecolor{brightpink}{rgb}{1.0,0.0,0.5}
\usepackage[colorlinks=true,linkcolor=black,citecolor=brightpink,urlcolor=black,
            bookmarksnumbered=true,bookmarksopen=true,pdfstartview=FitH]{hyperref}

\newcommand{\ii}{\mathrm{i}}
\newcommand{\ee}{\mathrm{e}}
\newcommand{\dd}{\mathrm{d}}
\newcommand{\Tr}{\operatorname{Tr}}
\newcommand{\PV}{\operatorname{PV}}
\newcommand{\Si}{\operatorname{Si}}
\newcommand{\Ci}{\operatorname{Ci}}
\newcommand{\cP}{\mathcal P}
\newcommand{\cM}{\mathcal M}
\newcommand{\cL}{\mathcal L}
\newcommand{\cH}{\mathcal H}
\newcommand{\cQ}{\mathcal Q}
\newcommand{\cC}{\mathcal C}
\newcommand{\cN}{\mathcal N}
\newcommand{\id}{\mathbb I}
\newcommand{\ord}{\mathcal O}
\newcommand{\ket}[1]{\lvert #1\rangle}
\newcommand{\bra}[1]{\langle #1\rvert}

\newtheorem*{theorem}{Theorem}

\hypersetup{pdftitle={Nash Equilibrium from Quantum-Field Entanglement},pdfauthor={Hao Xu}}

\begin{document}

\title{\Large\bfseries Nash Equilibrium from Quantum-Field Entanglement}
\author{Hao Xu\thanks{Corresponding author: \href{mailto:haoxu@yzu.edu.cn}{haoxu@yzu.edu.cn}}\\[6pt]
\small\textit{Center for Gravitation and Cosmology, College of Physical Science and Technology,}\\
\small\textit{Yangzhou University, Yangzhou 225002, China}}
\date{}
\maketitle

\begin{abstract}
Entanglement harvesting allows localized probes to extract entanglement from a quantum field, even when their interaction regions are spacelike separated. However, how the resource obtained by the probes can be operationalized remains largely open. In this work, we ask whether the harvested entanglement can influence the timing choices of observers acting as players. We construct a quantum input game for Alice and Bob, each controlling an Unruh-DeWitt detector that couples to the vacuum of a free massless scalar field in four-dimensional Minkowski spacetime. Each player can choose an early interaction time, a late interaction time, or exit. After joint entry, a referee sends independent question qubits, so that fixed local Bell measurements and a payment table convert the detector state into a payoff. The mean payoff is proportional to minus the expectation value of an entanglement witness and is nonpositive for every separable resource. We calculate the detector state to leading perturbative order and solve the participation game analytically. Away from payoff degeneracies, every timing equilibrium with positive payoff gives two trembling hand perfect participation equilibria, in addition to mutual exit. The total count can therefore be one, three, five, or seven. For one fixed switching function and timing menu, five selected delays give the sequence seven, three, one, seven, and one. The timing payoffs follow from the field's two-point functions, with the Hadamard and Pauli-Jordan contributions to the coherence competing against local excitation noise. The resulting equilibrium structure provides an operational setting for studying how vacuum correlations and causal propagation affect the strategic use of quantum-field entanglement.
\end{abstract}

\clearpage
\setcounter{tocdepth}{2}
\tableofcontents

\section{Introduction}

The Minkowski vacuum contains no particles for inertial observers, yet it possesses nontrivial quantum correlations and entanglement. The Reeh-Schlieder theorem reveals this structure through the cyclicity of the vacuum for the algebra of local operators associated with any open region~\cite{Reeh:1961ujh,Witten:2018zxz}. Suitable observables in spacelike separated regions can even violate Bell inequalities~\cite{SummersWerner1985,SummersWerner1987}. Operationally, localized probes provide a way to access these properties~\cite{Valentini1991}. Two initially uncorrelated probes can become entangled through local interactions with the vacuum, even when their interaction regions remain spacelike separated~\cite{Reznik2003,Reznik2005}. This process, known as entanglement harvesting~\cite{PozasKerstjens2015}, has been studied across a broad range of physical conditions~\cite{Salton2015,VerSteeg2009,PozasKerstjens2016,Ng2018,Henderson2019,CongMovingMirror2019,Cong2020,TjoaMann2020,HuZhangYu2022,JiZhangYu2024,Xu:2026efg,LiuBoundary2021}. These results establish that vacuum entanglement can be transferred to accessible quantum systems. However, while the quantum field theory can shape the entanglement one obtains, the question of how observers can use the entanglement acquired by their detectors as a resource remains largely unexplored. In this work, we take a further step \emph{from observers to players} and ask whether the harvested entanglement can serve as a strategic resource and how the underlying field correlations reorganize the Nash equilibria of the resulting game.

The strategic value of entanglement is not determined solely by the total amount of entanglement extracted. In a game, each player's payoff depends not only on their own strategy but also on the strategies of the other players. After multiple rounds of play, the players may reach an equilibrium state, known as a Nash equilibrium, in which no player can improve their payoff by changing their own strategy~\cite{Nash1950,Nash1951}. Classical correlated equilibria already show that shared advice can alter incentives~\cite{Aumann1974,Aumann1987}, while quantum games extend the available resources to entangled states and quantum operations~\cite{Meyer1999,Eisert1999,Khan2018}. Bell correlations can enhance equilibrium payoffs in games with aligned or conflicting interests~\cite{BrunnerLinden2013,Pappa2015}. In this work, we adopt the common model in entanglement harvesting, where Alice and Bob each possess an identical two-level detector interacting with the field. The difference is that we give Alice and Bob the option to choose whether to participate, and if so, whether to use an early or a late interaction window. Their choices determine the joint detector state that supplies the quantum resource for the game. We therefore study not only the resource they obtain, but also the stability of the choices that prepare it.

To make this question operational, we need a fixed rule that converts the detector state into a payoff. Entanglement alone does not imply Bell nonlocality~\cite{Horodecki2009,Brunner2014}, so a Bell test is not a general substitute for entanglement verification. This distinction also matters in collider applications~\cite{Abel2026}. We instead use the quantum input witness construction~\cite{Buscemi2012,Branciard2013}. After the detector interactions have ended, a referee sends question qubits to the two players. Each player measures the question jointly with their own detector and returns a classical bit. We introduce a prescribed phase calibration, local Bell readouts, and a fixed payment table to make the mean payoff proportional to minus an entanglement-witness expectation value. Every separable replacement gives a nonpositive payoff under the rules, even with arbitrary local readouts. A positive payoff thus has a definite resource interpretation.

With this verification rule fixed, the field determines which payoffs the different interaction times can produce. Microcausality requires local observables in spacelike separated regions to commute, while the quantum state can retain significant correlations across spacelike separations, and its Wightman function need not vanish there. Hadamard and Pauli-Jordan contributions enter the detector coherence~\cite{Tjoa2021,Zambianco2024}, where the former describes state correlations, and the latter carries the causal response. Together with local excitation noise, they determine the timing payoffs. For a free massless scalar field in four-dimensional Minkowski spacetime, the commutator is supported on the light cone. Compact interaction windows can therefore be strictly spacelike separated, intersect the propagating wavefront, or lie entirely beyond it. Varying the baseline delay between Alice's and Bob's timing allows us to compare these field-correlation regimes within the same protocol.

We calculate the detector state to leading perturbative order and solve the game analytically. Away from payoff degeneracies, each active timing equilibrium with positive reward gives two full equilibria, one with certain entry and one with entry mixed with exit. Together with mutual exit, this yields the complete count $N_{\rm THP}=1+2n_+\in\{1,3,5,7\}$, where $n_+$ counts the positive-reward active equilibria. All of these equilibria are trembling hand perfect. That is, they remain stable even if one player deviates infinitesimally from the equilibrium strategy. The other players need not to change their own strategies in response. In the massless field case, five selected delays within one fixed protocol yield $7\to3\to1\to7\to1$. These settings sample strictly spacelike interactions, the first light cone response lobe, carrier cancellation, the second lobe, and the post-wavefront Huygens regime. The two settings with seven equilibria correspond to anti-coordination and coordination in the active game, respectively. The construction thus connects field two-point functions to verified payoffs and, through the best-response conditions, to the complete equilibrium structure.

Our results show that vacuum entanglement in quantum field theory has operational significance beyond its extraction and detection by detectors. Even when the detector interactions are strictly spacelike separated, vacuum entanglement can support participation in a fixed quantum-input game. This allows observers to act as players, acquiring the resource through local couplings to the field and constructing strategies through the timing of their interactions. More broadly, quantum-field correlations can influence not only measurement outcomes but also the behavior that observers should adopt in a given quantum game. Moreover, since in principle all observers share the same physical system, namely our universe, this interconnection among observers may open up many new directions for future research.

The remainder of the paper is organized as follows. Section~\ref{sec2} derives the detector state and separates its Hadamard and Pauli-Jordan contributions. Section~\ref{sec3} constructs the quantum input game and proves the separable bound. Section~\ref{sec4} solves the timing and participation games, and establishes the complete equilibrium count. Section~\ref{sec5} evaluates the massless field response for a compactly supported switching function and examines five delay settings. Section~\ref{sec6} discusses the results and their extensions. Appendix~\ref{secA} considers one-sided changes of the local measurement, and Appendix~\ref{secB} proves trembling hand perfection.

\section{The entanglement harvesting model}\label{sec2}

We begin with a brief review of the entanglement harvesting model in four-dimensional Minkowski spacetime. The Unruh-DeWitt model describes a localized quantum system coupled to the field along its worldline. We consider two stationary two-level detectors coupled to a free massless scalar field. By tracing out the field after the interaction, we obtain the reduced density matrix of the detectors whose elements depend on the field correlations~\cite{Unruh1976,FewsterVerch2020}. This allows us to study what the probes detect without introducing a particle interpretation for the field.

\subsection{The reduced detector state}
Let Alice and Bob control detectors $A$ and $B$, respectively. Both detectors are stationary in a Minkowski inertial frame. The worldlines are
\begin{equation}
x_A(t)=(t,\bm0),\qquad x_B(t)=(t,\bm L),\qquad |\bm L|=L>0.
\end{equation}
Since the detectors are at rest, coordinate time is also their proper time. We choose Alice's switching function to be centered at $t=0$ and Bob's at $t=s$, so that
\begin{equation}
\chi_A(t)=\chi(t),\qquad \chi_B(t)=\chi(t-s).
\end{equation}
The switching function $\chi$ is even, sufficiently regular, and supported on $[-T,T]$, where $T>0$ is the interaction half-duration. Alice's detector therefore couples during $[-T,T]$, and Bob's during $[s-T,s+T]$. The delay $s$ specifies their relative schedule, and the two time variables in the interaction integrals remain independent. We set $\hbar=c=1$, so $T$, $L$, and $s$ have dimensions of length.

Each detector has ground state $\ket g$, excited state $\ket e$, energy gap $\Omega>0$ of inverse-length dimension, and a dimensionless weak coupling $\lambda$. The detector-field interaction is the standard monopole coupling~\cite{Unruh1976,PozasKerstjens2015},
\begin{align}
H_I(t)&=\lambda\sum_{\nu=A,B}\chi_\nu(t)\mu_\nu(t)\phi(x_\nu(t)),\\
\mu_\nu(t)&=\sigma_\nu^+\ee^{\ii\Omega t}+\sigma_\nu^-\ee^{-\ii\Omega t}.
\end{align}
Here $\nu\in\{A,B\}$, $\sigma_\nu^\pm$ raise and lower detector $\nu$, and $\phi(x_\nu(t))$ is the field evaluated on its worldline. The initial total state is $\rho_0:=\ket{gg}\bra{gg}\otimes\ket0\bra0$, where $\ket0$ is the field vacuum.

Now we can derive the leading-order reduced detector state by expanding the interaction picture evolution operator $U$ as
\begin{equation}
U=\id+U^{(1)}+U^{(2)}+\cdots,\quad
U^{(1)}=-\ii\int\dd t\,H_I(t),\quad
U^{(2)}=-\int\dd t\int_{-\infty}^{t}\dd t'\,H_I(t)H_I(t').
\end{equation}
The one-point function of the vacuum vanishes. Introducing field vectors
\begin{equation}
\ket{v_\nu}=\int\dd t\,\chi_\nu(t)\ee^{\ii\Omega t}\phi(x_\nu(t))\ket0,
\end{equation}
we have $U^{(1)}\ket{gg,0}=-\ii\lambda(\ket{eg}\ket{v_A}+\ket{ge}\ket{v_B})$. The term $U^{(1)}\rho_0U^{(1)\dagger}$ gives the one-excitation populations and their coherence,
\begin{equation}
\cP=\langle v_A|v_A\rangle=\langle v_B|v_B\rangle,
\qquad \cL(s)=\langle v_B|v_A\rangle.
\end{equation}
Translation invariance and identical switching profiles give the two detectors equal local responses. With
$W(t-t',L)=\langle0|\phi(t,\bm0)\phi(t',\bm L)|0\rangle$, we have
\begin{align}
\cP&=\int\dd t\,\dd t'\,\chi(t)\chi(t')
\ee^{-\ii\Omega(t-t')}W(t-t',0),
\label{eq8}
\\
\cL(s)&=\int\dd t\,\dd t'\,\chi(t)\chi(t'-s)
\ee^{\ii\Omega(t-t')}W(t'-t,L).
\label{eq9}
\end{align}

We write the time-ordered vacuum two-point function as $W_F(t,L):=\langle0|\mathcal T\phi(t,\bm0)\phi(0,\bm L)|0\rangle$, where $\mathcal T$ denotes time ordering, and the cross terms from $U^{(2)}\rho_0+\rho_0U^{(2)\dagger}$ give
\begin{equation}
\cM(s)=-\int\dd t\,\dd t'\,\chi(t)\chi(t'-s)\ee^{\ii\Omega(t+t')}W_F(t-t',L),
\label{eq10}
\end{equation}
In the ordered basis $(\ket{gg},\ket{ge},\ket{eg},\ket{ee})$ used in this work, the reduced state is
\begin{equation}
\rho_{AB}(s)=
\begin{pmatrix}
1-2\lambda^2\cP&0&0&\lambda^2\cM^*\\
0&\lambda^2\cP&\lambda^2\cL^*&0\\
0&\lambda^2\cL&\lambda^2\cP&0\\
\lambda^2\cM&0&0&0
\end{pmatrix}+\ord(\lambda^4),
\label{eq11}
\end{equation}
where equations~\eqref{eq8}, \eqref{eq9}, and \eqref{eq10} supply every coefficient in Eq.~\eqref{eq11}.

\subsection{Entanglement negativity}
Next, we calculate the amount of entanglement that these two detectors harvested from the quantum field theory. Write $\rho:=\rho_{AB}(s)$, the partial transposition on Bob's detector acts on product basis operators as
\begin{equation}
(\ket{i j}\bra{k l})^{T_B}=\ket{i l}\bra{k j}.
\end{equation}
Consequently, in the ordered basis $(\ket{gg},\ket{ge},\ket{eg},\ket{ee})$,
\begin{equation}
\rho^{T_B}=
\begin{pmatrix}
1-2\lambda^2\cP&0&0&\lambda^2\cL^*\\
0&\lambda^2\cP&\lambda^2\cM^*&0\\
0&\lambda^2\cM&\lambda^2\cP&0\\
\lambda^2\cL&0&0&0
\end{pmatrix}+\ord(\lambda^4).
\end{equation}
Writing $\cM=|\cM|\ee^{\ii\theta}$, the single-excitation block has eigenpairs
\begin{equation}
\frac{\ket{ge}\mp\ee^{\ii\theta}\ket{eg}}{\sqrt2},\qquad
\lambda^2(\cP\mp|\cM|)+\ord(\lambda^4).
\end{equation}
The negativity is defined by
$\cN(\rho)=(\|\rho^{T_B}\|_1-1)/2$, where
$\|X\|_1=\Tr\sqrt{X^\dagger X}$ is the trace norm. Since $\rho^{T_B}$ is Hermitian and has unit trace, $\cN$ equals the sum of the absolute values of its negative eigenvalues, in accordance with the standard definition of negativity~\cite{VidalWerner2002}. It is therefore positive exactly when the partial transpose has a negative eigenvalue. Applying the eigenvalues above gives
\begin{equation}
\cN(\rho)=\lambda^2(|\cM|-\cP)_++\ord(\lambda^4).
\end{equation}
Here $x_+=\max(x,0)$. The coherence $\cL$ does not enter the leading negativity because partial transposition moves it into the $(\ket{gg},\ket{ee})$ block, so that $\cL$-dependent lower eigenvalue starts only at order $\lambda^4$. Partial transposition on Alice's detector has the same spectrum because of the symmetry $\rho^{T_A}=(\rho^{T_B})^T$. For two qubits, positivity of the partial transpose is equivalent to the separability for the exact state~\cite{Peres1996,Horodecki1996}.

The leading entanglement condition compares the pair coherence $|\cM|$ with the local noise $\cP$. To understand what the field contributes to $\cM$, we first separate the phase due to the shift of Bob's switching function from the remaining field integral.
In Eq.~\eqref{eq10}, introducing the centered coordinates
\begin{equation}
z=\frac{t+t'-s}{2},\qquad \zeta=t-t'+s,
\end{equation}
the switching functions, detector phase, and field distribution become
\begin{equation}
\begin{gathered}
\chi(t)\chi(t'-s)=\chi(z+\zeta/2)\chi(z-\zeta/2),\\
\ee^{\ii\Omega(t+t')}=\ee^{\ii\Omega s}\ee^{2\ii\Omega z},\qquad
W_F(t-t',L)=W_F(\zeta-s,L).
\end{gathered}
\end{equation}
Thus the relative shift of the switching centers produces the factor $\ee^{\ii\Omega s}$ before any further integration. The support constraints $|t|\leq T$ and $|t'-s|\leq T$ imply
\begin{equation}
|\zeta|\leq2T,\qquad -T+|\zeta|/2\leq z\leq T-|\zeta|/2.
\end{equation}
The $z$ interval is symmetric. Since $\chi$ is even, the product of the two switches is even in $z$. The sine part of $\ee^{2\ii\Omega z}$ integrates to zero. Define
\begin{equation}
\cC_\Omega(\zeta)=\int_{-T+|\zeta|/2}^{T-|\zeta|/2}\dd z\,
\chi(z+\zeta/2)\chi(z-\zeta/2)\cos(2\Omega z)
\label{eq19}
\end{equation}
inside $|\zeta|\leq2T$, and zero outside. Then we have
\begin{equation}
\cM(s)=-\ee^{\ii\Omega s}\int\dd\zeta\,\cC_\Omega(\zeta)W_F(\zeta-s,L),\qquad
Z(s)=\int\dd\zeta\,\cC_\Omega(\zeta)W_F(\zeta-s,L).
\label{eq20}
\end{equation}
Thus $Z(s)=-\ee^{-\ii\Omega s}\cM(s)$ removes the known carrier phase and leaves $|Z|=|\cM|$, while the entanglement is unchanged.

We can now identify the real and imaginary parts of the remaining integral with two different field contributions. Writing the two-point function in terms of its symmetric and antisymmetric parts, the Wightman, Hadamard, and Pauli-Jordan distributions of the free real scalar field are defined by
\begin{align}
W(x,x')&=\langle0|\phi(x)\phi(x')|0\rangle\notag \\
G^{(1)}(x,x')&=\langle0|\{\phi(x),\phi(x')\}|0\rangle
=W(x,x')+W(x',x)\notag \\
[\phi(x),\phi(x')]&=\ii\Delta(x,x')\id,\qquad
\ii\Delta(x,x')=W(x,x')-W(x',x).
\end{align}
For a Hermitian field, $W(x',x)=W(x,x')^*$, so $G^{(1)}=2\operatorname{Re}W$ and $\Delta=2\operatorname{Im}W$.  The massless mode expansion with $[a_{\bm k},a_{\bm k'}^\dagger]=\delta^3(\bm k-\bm k')$ gives
\begin{align}
W(t,L)&=\int\frac{\dd^3k}{2(2\pi)^3|\bm k|}\,
\ee^{-\ii|\bm k|t+\ii\bm k\cdot\bm L}\nonumber\\
&=\frac1{4\pi^2L}\lim_{\epsilon\to0^+}\int_0^\infty\dd k\,
\sin(kL)\ee^{-(\epsilon+\ii t)k}
=\frac1{4\pi^2}\frac1{L^2-(t-\ii0)^2}.
\end{align}
Using the Sokhotski-Plemel formula
\begin{equation}
\frac1{x\pm\ii0}=\PV\frac1x\mp\ii\pi\delta(x)
\end{equation}
in the partial fractions at $t=\pm L$, one obtains
\begin{align}
G^{(1)}(t,L)&=W(t,L)+W(-t,L)
=\frac1{2\pi^2}\PV\frac1{L^2-t^2},  \\
\Delta(t,L)&=\frac{W(t,L)-W(-t,L)}{\ii}
=-\frac{\delta(t-L)-\delta(t+L)}{4\pi L}.
\label{eq25}
\end{align}
Here principal value $\PV$ means symmetric exclusion around each simple pole before taking the exclusion radius to zero. Then the time ordering is a separate operation:
\begin{align}
W_F(t,L)&=\Theta(t)W(t,L)+\Theta(-t)W(-t,L)\nonumber\\
&=\frac12G^{(1)}(t,L)+\frac{\ii}{2}\operatorname{sgn}(t)\Delta(t,L)\nonumber\\
&=\frac1{4\pi^2}\PV\frac1{L^2-t^2}
-\frac{\ii}{8\pi L}[\delta(t-L)+\delta(t+L)].
\label{eq26}
\end{align}
Here $\Theta$ is the Heaviside step function. Writing the carrier-removed coherence as $Z(s)=\cH(s)+\ii\cQ(s)$, Eqs.~\eqref{eq20} and~\eqref{eq26} give
\begin{align}
\cH(s)&=\frac1{4\pi^2}\PV\int\frac{\cC_\Omega(\zeta)}{L^2-(\zeta-s)^2}\dd\zeta,\\
\cQ(s)&=-\frac{\cC_\Omega(s-L)+\cC_\Omega(s+L)}{8\pi L}.
\label{eq28}
\end{align}
These terms separate the anticommutator correlations from the commutator contribution~\cite{Tjoa2021,Zambianco2024}. For our identical even switching profiles, they form two real quadratures, so $|Z|^2=\cH^2+\cQ^2$.

We can now see how the causal relation between the interaction regions enters the detector state. The Hadamard contribution $\cH$ depends on the field state, while $\cQ$ comes from the state-independent commutator fixed by the Klein-Gordon equation. When the regions are strictly spacelike separated, microcausality gives $\cQ=0$, but $\cH$ can remain nonzero. Any positive spacelike payoff must therefore come from the vacuum correlations. If the regions intersect the light cone shell, the pair coherence may also contain a causal contribution. This concerns the propagation of information between detectors, and such propagation need not be accompanied by net energy exchange~\cite{Jonsson2015}.

\section{A quantum input game}\label{sec3}

Having obtained the detector state, we need to specify how the players will use it. The field interactions end before verification begins, and the referee then acts only on the detectors and the question qubits. We need to construct a set of rules such that entanglement can serve as a benchmark for deciding how to play this game.

\subsection{The entanglement witness}
Firstly we construct the observable whose expectation value will set the payoff. The eigenvector associated with the potentially negative eigenvalue of $\rho^{T_B}$ gives a convenient witness. Let $\theta=\arg\cM$ and $\ket{\eta_\theta}=(\ket{ge}-\ee^{\ii\theta}\ket{eg})/\sqrt2$. Taking the partial transpose of its projector gives
\begin{equation}
W_\theta=(\ket{\eta_\theta}\bra{\eta_\theta})^{T_B}
=\frac12\begin{pmatrix}
0&0&0&-\ee^{-\ii\theta}\\
0&1&0&0\\
0&0&1&0\\
-\ee^{\ii\theta}&0&0&0
\end{pmatrix}.
\label{eq29}
\end{equation}
Although $\ket{\eta_\theta}$ has support only in the single-excitation sector, partial transposition moves its coherences to the sector spanned by $\ket{gg},\ket{ee}$. It is not a unitary change of basis or a physical state transformation. For arbitrary positive operators $A,B$,
\begin{equation}
\Tr[W_\theta(A\otimes B)]
=\bra{\eta_\theta}A\otimes B^T\ket{\eta_\theta}\geq0,
\label{eq30}
\end{equation}
because transposition preserves positive semidefiniteness. Thus $W_\theta$ is nonnegative on all separable states, although it has a negative eigenvalue itself, as is permitted for an entanglement witness~\cite{GuhneToth2009}. For the detector state,
\begin{align}
\Tr(W_\theta\rho)=\lambda^2(\cP-|\cM|)+\ord(\lambda^4).
\end{align}
This witness tests the leading negative partial transpose direction of the detector state through a linear expectation value. Its relation to the negativity follows from the form of our detector state and does not hold for an arbitrary two-qubit state. We write $W_0:=W_{\theta=0}$ for the real witness.

The coherence phase changes when a switching function is translated. If we want to use the same real witness $W_0$ for every history, we must account for this phase before measuring. In the energy basis used above, the required phase is
\begin{equation}
\theta(s)=\arg\cM(s)=\Omega s+\arg Z(s)+\pi\pmod{2\pi}.
\label{eq32}
\end{equation}
Applying $U_s=\operatorname{diag}(1,\ee^{-\ii\theta(s)})$ to Alice's detector, we can define
\begin{equation}
\widetilde\rho_{AB}=(U_s\otimes\id)\rho(U_s^\dagger\otimes\id).
\end{equation}
Its $ee,gg$ element is $\lambda^2|\cM|+\ord(\lambda^4)$, and
\begin{equation}
(U_s^\dagger\otimes\id)W_0(U_s\otimes\id)=W_\theta,
\qquad \Tr(W_0\widetilde\rho_{AB})=\Tr(W_\theta\rho).
\end{equation}
The prescribed rotation aligns the measured quadrature with $\cM$, allowing the same real witness $W_0$ to test the coherence magnitude for each history.

Equation~\eqref{eq32} assumes centers $(0,s)$. For general centers $(a,b)$, set $s=b-a$ and translate both integration times in the pair coherence integral by $a$. The vacuum two-point function and time ordering depend only on time differences, whereas $\ee^{\ii\Omega(t+t')}$ gains $\ee^{2\ii\Omega a}$. Consequently,
\begin{equation}
\cM_{a,b}=\ee^{2\ii\Omega a}\cM(b-a)
=-\ee^{\ii\Omega(a+b)}Z(b-a).
\end{equation}
The coherence phase therefore increases by $2\Omega a$ relative to $\theta(b-a)$, modulo $2\pi$, so the relative delay alone does not fix the required calibration. To state the calibration for the timing menu analyzed in Sec.~\ref{sec4}, let $i,j\in\{0,1\}$ denote Alice's and Bob's early/late choices, let $\delta>0$ be the separation between the two centers in either player's menu, and let $\tau$ be Bob's baseline delay relative to Alice. Alice's and Bob's switching centers are $i\delta$ and $\tau+j\delta$, respectively, so their relative delay is $s_{ij}:=\tau+(j-i)\delta$. The prescribed rotation is
\begin{equation}
U_{ij}=\operatorname{diag}(1,\ee^{-\ii\theta_{ij}}),\quad
\theta_{ij}=\Omega[\tau+(i+j)\delta]+\arg Z(s_{ij})+\pi
=\theta(s_{ij})+2\Omega i\delta\pmod{2\pi}.
\end{equation}
It gives $\ee^{-\ii\theta_{ij}}\cM_{i\delta,\tau+j\delta}=|Z(s_{ij})|$, so the calibrated reward depends only on the relative delay.

This correction is part of the prescribed apparatus. A trusted controller reads the registered switching centers and applies $U_{ij}$ after the field interaction but before the referee prepares the questions. The phase for each history is fixed in advance by the detector-field model or by an independent calibration, rather than estimated from that round's answers. For spacelike interactions, the records can be compared later in a common causal future. Since the question labels have not yet been drawn, this comparison carries no question dependent information. The players choose their detectors' interaction times, while the calibration follows the registered choices.

\subsection{Quantum questions and Bell measurements}

We use this calibrated witness to realize a mean payoff
\begin{equation}
G=-\Gamma\Tr(W_0\widetilde\rho_{AB}),\qquad \Gamma>0,
\label{eq37}
\end{equation}
using only the question labels and the players' classical answers. We must therefore express this expectation value as a linear combination of observable answer probabilities. Table~\ref{tab1} summarizes the resulting protocol, and the payment rule we derive below.

\begin{table}[H]
\centering
\caption{One round of the fixed protocol. All strategic choices precede state preparation and verification.}
\label{tab1}
\begin{tabular}{p{.15\linewidth}p{.76\linewidth}}
\toprule
Stage & Operation\\
\midrule
Commit & Each player chooses exit, early entry, or late entry.\\
Prepare & If both enter, their detectors interact locally with the vacuum at the committed switching centers.\\
Calibrate & A controller applies the prescribed phase correction determined by the timing history.\\
Verify & The referee distributes independent quantum questions. Each player performs the fixed local binary Bell readout and returns one classical bit without communication.\\
Pay & The referee applies the same payment table to every timing history. A lone entrant loses the unit stake.\\
\bottomrule
\end{tabular}
\end{table}
We choose the question states from the four Bloch vectors
\begin{equation}
\bm r_0=\frac{(1,1,1)}{\sqrt3},\quad
\bm r_1=\frac{(1,-1,-1)}{\sqrt3},\quad
\bm r_2=\frac{(-1,1,-1)}{\sqrt3},\quad
\bm r_3=\frac{(-1,-1,1)}{\sqrt3}.
\label{eq38}
\end{equation}
Here $\bm r_q=(r_{qx},r_{qy},r_{qz})$, and similarly $\bm r_r=(r_{rx},r_{ry},r_{rz})$ for Bob's question label $r$.
Define $\tau_q=(\id+\bm r_q\cdot\bm\sigma)/2$, where $\bm\sigma=(\sigma_x,\sigma_y,\sigma_z)$ is the Pauli matrices. We have
\begin{equation}
\tau_q=\frac12\begin{pmatrix}1+r_{qz}&r_{qx}-\ii r_{qy}\\r_{qx}+\ii r_{qy}&1-r_{qz}\end{pmatrix}.
\end{equation}
The overlaps are $\Tr(\tau_q\tau_r)=1$ for $q=r$ and $1/3$ otherwise. Hence we can define
\begin{equation}
D_q=\frac{3\tau_q-\id}{2}=\frac{\id+3\bm r_q\cdot\bm\sigma}{4},\qquad
\Tr(D_q\tau_r)=\delta_{qr}.
\end{equation}
The four states are linearly independent in the four-dimensional real vector space of Hermitian $2\times2$ matrices. Their sixteen tensor products span the sixteen-dimensional Hermitian two-qubit operator space. Taking $W_0$ from Eq.~\eqref{eq29} at $\theta=0$, the dual-basis expansion and direct evaluation of the matrix traces give
\begin{align}
W_0&=\sum_{q,r=0}^{3}\beta_{qr}\tau_q^T\otimes\tau_r^T,\\
\beta_{qr}&=\Tr[(D_q^T\otimes D_r^T)W_0]
=\frac{1-9r_{qx}r_{rx}+9r_{qy}r_{ry}-9r_{qz}r_{rz}}{16}.
\label{eq42}
\end{align}
Substitution of Eq.~\eqref{eq38} yields
\begin{equation}
(\beta_{qr})_{q,r=0}^{3}=\frac18\begin{pmatrix}-1&-1&5&-1\\-1&-1&-1&5\\5&-1&-1&-1\\-1&5&-1&-1\end{pmatrix}.
\label{eq43}
\end{equation}

The referee sends one question qubit to each laboratory. Alice now holds $Q_A$ and her detector $A$. Her binary readout distinguishes the Bell state from its orthogonal complement, where
\begin{equation}
\ket{\Phi^+}=\frac{\ket{gg}+\ket{ee}}{\sqrt2},\qquad
\Phi^+:=\ket{\Phi^+}\bra{\Phi^+}.
\end{equation}
We write $\Phi^+_{Q_AA}$ for the projector on $Q_A\otimes A$ and $\Phi^+_{BQ_B}$ for the projector on $B\otimes Q_B$. The subscripts specify the qubits and their tensor order. Each projector acts on a question qubit and the detector in the same laboratory.
She reports $1$ for the projection onto $\ket{\Phi^+}$ and $0$ otherwise. Bob does the same locally. For arbitrary $2\times2$ matrices $R,S$, we have
\begin{align}
\bra{\Phi^+}R\otimes S\ket{\Phi^+}
=\frac12\sum_{m,n=0}^{1}\bra mR\ket n\bra mS\ket n
=\frac12\sum_{m,n}R_{mn}S_{mn}
=\frac12\Tr(R^TS).
\label{eq45}
\end{align}
Applying this identity to each detector factor in a linear expansion of $\widetilde\rho_{AB}$ gives the probability of both player reports $1$ as
\begin{equation}
p(1,1|q,r)=\Tr[(\Phi^+_{Q_AA}\otimes\Phi^+_{BQ_B})
(\tau_q^{Q_A}\otimes\widetilde\rho_{AB}\otimes\tau_r^{Q_B})]
=\frac14\Tr[(\tau_q^T\otimes\tau_r^T)\widetilde\rho_{AB}].
\label{eq46}
\end{equation}
The tensor factor ordering in the trace is $Q_A,A,B,Q_B$.
For an entangled resource, the joint probability need not factorize into the two local marginals.

We can now express the target payoff in terms of these probabilities. Substituting the expansion of $W_0$ from Eq.~\eqref{eq42} into Eq.~\eqref{eq37}, and using Eq.~\eqref{eq46}, gives
\begin{align}
-\Gamma\Tr(W_0\widetilde\rho_{AB})
=-\Gamma\sum_{q,r}\beta_{qr}
\Tr[(\tau_q^T\otimes\tau_r^T)\widetilde\rho_{AB}]
=-4\Gamma\sum_{q,r}\beta_{qr}p(1,1|q,r),
\label{eq47}
\end{align}
The factor of four compensates for the two local Bell contractions, each of which contributes a factor of $1/2$. It remains to choose a single-round payment whose average reproduces this weighted sum.

\subsection{Payoffs and the separable bound}
The referee draws each label independently and uniformly from $\{0,1,2,3\}$, so each question pair occurs with probability $1/16$. The labels are not revealed to the players, who receive only the question qubits and return binary answers $a,b$. Let $r(q,r,a,b)$ be the single-round payment to each player. We assign zero unless both answers are $1$, so the mean payoff is
\begin{align}
G=\frac1{16}\sum_{q,r,a,b}r(q,r,a,b)p(a,b|q,r)
=\frac1{16}\sum_{q,r}r(q,r,1,1)p(1,1|q,r).
\end{align}
Comparing this expression with Eq.~\eqref{eq47}, we can obtain the target by choosing
\begin{equation}
r(q,r,a,b)=-64\Gamma\beta_{qr}\delta_{a1}\delta_{b1}.
\end{equation}
Substituting the coefficients in Eq.~\eqref{eq43} gives the double-click payment table
\begin{equation}
r(q,r,1,1)=\Gamma
\begin{pmatrix}
8&8&-40&8\\
8&8&8&-40\\
-40&8&8&8\\
8&-40&8&8
\end{pmatrix}_{qr},
\qquad \Gamma>0,
\end{equation}
Every round is included in the average, including those with zero. The resulting protocol therefore gives
\begin{equation}
G=-\Gamma\Tr(W_0\widetilde\rho_{AB})
=\Gamma\lambda^2(|Z|-\cP)+\ord(\lambda^4).
\label{eq51}
\end{equation}
For the timing game, we define the leading payoff as $g(s):=\Gamma\lambda^2[\sqrt{\cH(s)^2+\cQ(s)^2}-\cP]$, so that $G(s)=g(s)+\ord(\lambda^4)$.
The equality between the average payment and the witness expectation is exact for the prescribed Bell measurements.  Different payment tables could give different games and different equilibria. We choose this table because its positive payoff certifies the leading negative partial transpose entanglement of the detector state, while every separable replacement gives a nonpositive payoff under the same verification conditions. The choice is fixed before the timing histories are compared. Once it has been made, the players choose only whether to enter and when their detectors interact with the field. The field correlations then determine the payoffs available for those choices.

We now allow arbitrary local binary measurements on a separable state and show that the average payoff is still nonpositive. This part of the argument is exact and does not use the perturbative detector state. For this comparison, let the detector state be separable,
\begin{equation}
\rho_{AB}=\sum_k p_k\rho_A^{(k)}\otimes\rho_B^{(k)},
\qquad p_k\geq0,
\qquad \sum_kp_k=1.
\label{eq52}
\end{equation}
The label $k$ is a classical shared branch. Local ancillary systems and their fixed states can be absorbed into $\rho_A^{(k)}$ and $\rho_B^{(k)}$.

Alice may use any answer-$1$ effect $0\leq A_1\leq\id_{Q_AA}$, and Bob may use any $0\leq B_1\leq\id_{BQ_B}$. For a fixed branch $k$, Alice's local probability is
\begin{align}
p_A(1|q,k)
&=\Tr_{Q_AA}[A_1(\tau_q\otimes\rho_A^{(k)})]\nonumber\\
&=\Tr_{Q_AA}\!\left[(\id\otimes\sqrt{\rho_A^{(k)}})A_1
(\id\otimes\sqrt{\rho_A^{(k)}})(\tau_q\otimes\id_A)\right]\nonumber\\
&=\Tr_{Q_A}(E_A^{(k)}\tau_q),
\end{align}
where cyclicity of the full trace was used in the second line, and
\begin{equation}
E_A^{(k)}=
\Tr_A\!\left[(\id\otimes\sqrt{\rho_A^{(k)}})A_1
(\id\otimes\sqrt{\rho_A^{(k)}})\right].
\end{equation}
The operator $E_A^{(k)}$ acts on the question space. It combines Alice's fixed detector state in branch $k$ with her joint measurement, so that her answer probability can be computed from the question qubit alone. Similarly, on Bob's side we obtain
\begin{equation}
E_B^{(k)}=
\Tr_B\!\left[(\sqrt{\rho_B^{(k)}}\otimes\id_{Q_B})B_1
(\sqrt{\rho_B^{(k)}}\otimes\id_{Q_B})\right].
\end{equation}

These operators are valid effects on the two-dimensional question spaces. Congruence by a positive square root and partial trace preserve positivity, so $E_A^{(k)},E_B^{(k)}\geq0$. Moreover,
\begin{equation}
E_A^{(k)}\leq
\Tr_A(\id_{Q_A}\otimes\rho_A^{(k)})=\id_{Q_A},
\qquad
E_B^{(k)}\leq\id_{Q_B}.
\end{equation}

Within branch $k$, both the resource and the laboratory measurements factor across Alice and Bob. The rewarded joint probability therefore factorizes as
\begin{align}
p(1,1|q,r,k)
=p_A(1|q,k)p_B(1|r,k)=\Tr(E_A^{(k)}\tau_q)\Tr(E_B^{(k)}\tau_r).
\label{eq57}
\end{align}
The observable probability is the convex mixture
\begin{equation}
p(1,1|q,r)=\sum_kp_k\,p(1,1|q,r,k).
\end{equation}
This factorization is the step that fails for an entangled resource.

To combine Eq.~\eqref{eq57} with the operator expansion of $W_0$, we use
\begin{equation}
\Tr(E\tau)=\Tr[(E\tau)^T]=\Tr(E^T\tau^T),
\end{equation}
the weighted probability sum then becomes
\begin{align}
\sum_{q,r}\beta_{qr}p(1,1|q,r)
&=\sum_k p_k\sum_{q,r}\beta_{qr}
\Tr[(E_A^{(k)})^T\tau_q^T]\Tr[(E_B^{(k)})^T\tau_r^T]\notag\\
&=\sum_kp_k\Tr[W_0((E_A^{(k)})^T\otimes(E_B^{(k)})^T)]\geq0,
\label{eq60}
\end{align}
where the expansion in Eq.~\eqref{eq42} was used in the equality. Transposition preserves positivity, and Eq.~\eqref{eq30} makes every term in the final sum nonnegative. Hence
\begin{equation}
G_{\rm sep}=-4\Gamma\sum_{q,r}\beta_{qr}p(1,1|q,r)\leq0.
\end{equation}

The connection with the prescribed Bell readout is explicit. If $A_1=\Phi^+_{Q_AA}$, Eq.~\eqref{eq45} gives, for every question state,
\begin{equation}
\Tr(E_A^{(k)}\tau_q)
=\frac12\Tr(\tau_q^T\rho_A^{(k)})
=\Tr\!\left[\frac{(\rho_A^{(k)})^T}{2}\tau_q\right].
\end{equation}
Therefore
\begin{equation}
E_A^{(k)}=\frac{(\rho_A^{(k)})^T}{2},
\qquad
E_B^{(k)}=\frac{(\rho_B^{(k)})^T}{2}
\end{equation}
when both players use Bell effects. Equation~\eqref{eq60} then reduces to
\begin{equation}
\sum_{q,r}\beta_{qr}p(1,1|q,r)=\frac14\Tr(W_0\rho_{AB})\geq0.
\end{equation}
For an entangled calibrated resource $\widetilde\rho_{AB}$, no decomposition of the form in Eq.~\eqref{eq52} exists. The fixed Bell readout gives $G=-\Gamma\Tr(W_0\widetilde\rho_{AB})>0$ whenever the witness expectation is negative. The positive payoff is therefore inaccessible to all separable resources under arbitrary local readouts, not merely to separable resources using Bell readouts.

The separable decomposition in Eq.~\eqref{eq52} already allows the players to share classical randomness and coordinate their local operations. Even with this freedom and arbitrary local
readouts, a separable resource cannot give a positive payoff. The best payoff is zero, attained when both players always report $0$. This comparison allows more freedom than the timing game itself, where the Bell readouts remain fixed. Appendix~\ref{secA} separately examines whether changing only one player's measurement can improve the payoff.

\section{Counting Nash equilibria}\label{sec4}

In this section, we count the equilibria arising from fixed and probabilistic timing choices. We first identify the pure and mixed timing equilibria under joint entry, then extend every positive-reward equilibrium to include entry and exit. This gives the complete count $N_{\rm THP}$, with the proof of trembling hand perfection given in Appendix~\ref{secB}.

\subsection{Timing equilibria}
We first identify the strategies available to the players when they are certain to participate in the game. Each then chooses timing $0$ or $1$, meaning an early or a late interaction. The separation between a player's two windows is $\delta>0$, and Bob's menu is delayed by $\tau$ relative to Alice's. Their timing choices $i,j\in\{0,1\}$ give switching centers $i\delta$ and $\tau+j\delta$, respectively, with relative delay
\begin{equation}
s_{ij}=\tau+(j-i)\delta.
\end{equation}
After the phase calibration, the active game has common matrix
\begin{equation}
\begin{pmatrix}g_0&g_+\\g_-&g_0\end{pmatrix},\qquad
g_0=g(\tau),\quad g_\pm=g(\tau\pm\delta).
\end{equation}
For the rest of this section $g_0,g_+,g_-$ are arbitrary real numbers and the result is exact for a finite game. In the quantum field application they are the leading-order payoffs.

A Nash equilibrium is a pair of strategies for which neither player can improve their payoff through a one-sided deviation~\cite{Nash1951}. Assume $g_0\ne g_\pm$ throughout the timing classification. We begin with fixed choices, or pure strategies. At $(i,j)=(0,0)$, Alice could obtain $g_-$ by switching to late, while Bob could obtain $g_+$. Thus this profile is an equilibrium exactly when $g_0>\max(g_-,g_+)$. Checking the other three action pairs gives
\begin{center}
\begin{tabular}{p{.32\linewidth}p{.29\linewidth}p{.23\linewidth}}
\toprule Condition & Pure equilibria $(i,j)$ & Common rewards\\\midrule
$g_0>\max(g_-,g_+)$& $(0,0),(1,1)$ & $g_0,g_0$\\
$g_0<\min(g_-,g_+)$& $(0,1),(1,0)$ & $g_+,g_-$\\
$g_-<g_0<g_+$& $(0,1)$ & $g_+$\\
$g_+<g_0<g_-$& $(1,0)$ & $g_-$\\
\bottomrule
\end{tabular}
\end{center}
The first region gives coordination, the second anti-coordination, and the last two a single directional equilibrium.

We next ask whether the players can also be in equilibrium while choosing early and late with nonzero probabilities. Let $x$ be Alice's probability of timing $0$, and $y$ Bob's. Alice's conditional payoffs for the two pure timings are
\begin{equation}
A_0(y)=yg_0+(1-y)g_+,\qquad A_1(y)=yg_-+(1-y)g_0.
\end{equation}
Bob's are
\begin{equation}
B_0(x)=xg_0+(1-x)g_-,\qquad B_1(x)=xg_++(1-x)g_0.
\end{equation}
For both timings to be used in equilibrium, the supported actions must have equal payoffs, so $A_0(y)=A_1(y)$ and $B_0(x)=B_1(x)$. Solving these indifference conditions gives
\begin{equation}
x_* =\frac{g_0-g_-}{2g_0-g_--g_+},\qquad
y_* =\frac{g_0-g_+}{2g_0-g_--g_+},\qquad
u_* =\frac{g_0^2-g_-g_+}{2g_0-g_--g_+}.
\end{equation}

These solutions describe a genuine mixed equilibrium only if $0<x_*,y_*<1$. This holds precisely when $g_0$ is either the maximum or the minimum of the three rewards. There is then one additional timing equilibrium, namely the probability pair $(x_*,y_*)$. Each such pair counts as one equilibrium, rather than a separate equilibrium for each realized timing outcome. If $g_0$ lies between the off-diagonal rewards, no interior mixed equilibrium exists. The case $2g_0=g_-+g_+$, where the fractions are undefined, belongs to this directional region. Nor can exactly one player mix under $g_0\ne g_\pm$, since indifference against a pure timing would require an excluded equality. The active game therefore has three equilibria in the coordination and anti-coordination regions, and one in either directional region.

Existence alone is not enough once exit becomes available. We must also check the rewards of the pure equilibria listed above and, when present, the mixed reward $u_*$. Let $n_+$ count all active equilibria with strictly positive common reward, including both pure and mixed strategies. The ordering of $g_0,g_+,g_-$ determines which equilibria exist, where each equilibrium's reward sign determines whether it can support participation. Table~\ref{tab2} collects this positive-reward count.

\begin{table}[H]
\centering
\caption{Exact count of positive reward active equilibria. Here $\mathbf 1[C]$ is one when $C$ holds and zero otherwise. Equalities $g_0=g_\pm$ and zero active equilibrium rewards are excluded.}
\label{tab2}
\begin{tabular}{p{.42\linewidth}p{.47\linewidth}}
\toprule
Active reward region & $n_+$\\
\midrule
$g_0>\max\{g_+,g_-\}$ & $2\mathbf 1[g_0>0]+\mathbf 1[g_0^2>g_+g_-]$\\
$g_0<\min\{g_+,g_-\}$ & $\mathbf 1[g_+>0]+\mathbf 1[g_->0]+\mathbf 1[g_0^2<g_+g_-]$\\
$g_-<g_0<g_+$ & $\mathbf 1[g_+>0]$\\
$g_+<g_0<g_-$ & $\mathbf 1[g_->0]$\\
\bottomrule
\end{tabular}
\end{table}

\subsection{Participation equilibria}

We now set exit as a third action $E$, giving each player the menu $\{E,0,1\}$. Exit always pays zero, while entering alone costs one. With row and column order $(E,0,1)$, the full payoff matrices for Alice and Bob are, respectively,
\begin{equation}
\begin{pmatrix}0&0&0\\-1&g_0&g_+\\-1&g_-&g_0\end{pmatrix},\qquad
\begin{pmatrix}0&-1&-1\\0&g_0&g_+\\0&g_-&g_0\end{pmatrix}.
\label{eq70}
\end{equation}

Let $\alpha,\beta$ be Alice's and Bob's probabilities of entering. If $\alpha=0$, Bob faces certain exit, so entering loses one and $\beta=0$. Similarly with the players exchanged. Thus every nonexit equilibrium has $\alpha,\beta>0$.

Because entering alone incurs the same loss for either timing, every equilibrium with entry must retain an active timing equilibrium conditional on participation.

We can therefore start from the active equilibria already found and determine how often the players enter. Let $u$ be the common payoff of one such equilibrium. Keeping its conditional timing distribution, Alice's expected reward from entry is
\begin{equation}
\beta u-(1-\beta)=\beta(1+u)-1,
\label{eq71}
\end{equation}
whereas exit gives zero. Bob's entry reward is similarly $\alpha(1+u)-1$.

For $u<0$, entry has negative expected payoff for every opponent entry probability, while exit gives zero and is optimal. For $u>0$, certain entry by one player makes entry optimal for the other. Thus both enter with certainty, or both mix entry with exit. In the latter case, equating the entry payoff in Eq.~\eqref{eq71} to zero gives
\begin{equation}
\alpha=\beta=\frac1{1+u}.
\end{equation}
These probabilities lie strictly between zero and one precisely when $u>0$. The two participation equilibria corresponding to the active profile ($x,y$) are
\begin{align}
&\bigl((0,x,1-x),(0,y,1-y)\bigr),\\
&\bigl((1-\alpha,\alpha x,\alpha(1-x)),
       (1-\alpha,\alpha y,\alpha(1-y))\bigr),\qquad \alpha=(1+u)^{-1}.
\end{align}
The entries of each vector are ordered $E,0,1$. Each of the $n_+$ positive-reward timing equilibria therefore gives two distinct participation equilibria. This applies whether its timing choices are pure or mixed. Together with mutual exit, these exhaust the Nash equilibria away from $u=0$. These Nash equilibria are also trembling hand perfect, because their supported actions remain best responses against suitable completely mixed profiles approaching equilibrium~\cite{Selten1975}. Appendix~\ref{secB} gives the explicit perturbations for all cases.

\begin{samepage}
Now we have the following result.
\begin{theorem}[Complete finite game count]
For Eq.~\eqref{eq70}, assume $g_0\ne g_\pm$ and every active equilibrium has nonzero reward. Let $n_+$ count the positive-reward active equilibria, including both pure and mixed timings. All Nash equilibria of the full game are trembling hand perfect, and their number is
\begin{equation}
N_{\rm THP}=1+2n_+\in\{1,3,5,7\}.
\label{eq75}
\end{equation}
\end{theorem}
\end{samepage}

The number five is algebraically possible when the active game has exactly two positive equilibria. For example,
\begin{equation}
\begin{pmatrix}1&-3\\-3&1\end{pmatrix}
\end{equation}
has two pure coordinated equilibria of reward $1$, and an equal mixing equilibrium of reward $-1$. Hence $n_+=2$ and $N_{\rm THP}=5$.

\section{Results for a massless scalar field}\label{sec5}

In this section, we use a concrete model to illustrate the conclusions obtained above. Since we consider the free massless scalar field in four-dimensional Minkowski spacetime, and its commutator propagates only on the null cone, this allows us to rigorously distinguish strictly spacelike histories from histories that intersect the wavefront and from histories that lie entirely after it. Other field theories, such as massive fields, may lead to some differences from our results, but the analysis can be extended naturally.

\subsection{Causal geometry of the interactions}
For $L>2T$, the maximum magnitude of the time difference over the supports of the two switching functions is $|s|+2T$, so all event pairs are strictly spacelike if
\begin{equation}
|s|+2T<L.
\end{equation}
For $s\geq0$, the three regimes are
\begin{align}
0\leq s<L-2T &: \text{all event pairs strictly spacelike},\\
L-2T<s<L+2T &: \text{the supports intersect the light cone shell},\\
s>L+2T &: \text{all event pairs strictly timelike}.
\end{align}
The boundaries are limiting contacts of compact supports. The switching functions may vanish there, but these contacts do not satisfy the strict spacelike inequalities. Because the massless commutator in Eq.~\eqref{eq25} lives only on $t=\pm L$, the convolution $\cQ$ vanishes in the first and last open regimes. This is Huygens propagation in three spatial dimensions. The Hadamard distribution does not have this restricted support.

Introducing the dimensionless separation $\ell:=L/T$, at baseline $\tau=0$, the four timing histories have relative delays $s=0,0,+\delta,-\delta$. All four are strictly spacelike precisely when
\begin{equation}
\ell>2,\qquad 0<\frac{\delta}{T}<\ell-2.
\end{equation}
At baseline $\tau=2L$, the earliest sampled delay is $2L-\delta$. The same inequality ensures $2L-\delta>L+2T$, so all four histories are strictly beyond the light cone shell. These two conditions can be enforced simultaneously before we consider payoff signs.

To make both the causal geometry and the response integrals explicit, we choose the compactly supported switching function
\begin{equation}
\chi(t)=\begin{cases}\cos^2(\pi t/2T),&|t|\leq T,\\0,&|t|>T.\end{cases}
\label{eq82}
\end{equation}
The squared cosine is nonnegative, and both its value and first derivative vanish at the boundary. Its Fourier transform therefore decays as the inverse cube of frequency. A cosine cut off at its zeros could also be considered, but it has less endpoint regularity and gives a different response. The present choice keeps the leading integrals finite and allows simple analytic formulas.

With the convention $\widetilde\chi(\omega)=\int\dd t\,\chi(t)\ee^{\ii\omega t}$, we have
\begin{align}
\widetilde\chi(\omega)
=T\frac{\pi^2\sin(\omega T)}{(\omega T)[\pi^2-(\omega T)^2]},
\label{eq83}
\end{align}
and the apparent singularities at $\omega=0,\pm\pi/T$ are removable.

\subsection{Analytic response functions}
Set the dimensionless convolution variable $x:=\zeta/T$ and relative delay $y:=s/T$. We choose $\Omega T=3\pi/2$ because it substantially simplifies the convolution and the final analytic expressions. Other values of $\Omega T$ can be treated by the same method, but require more extensive calculations. Substituting the switching function in Eq.~\eqref{eq82} into Eq.~\eqref{eq19} and performing the $z$ integral gives the normalized kernel $c(x):=\cC_\Omega(Tx)/T$ as
\begin{equation}
c(x)=\frac{\sin^5(\pi|x|/2)}{15\pi}\quad (|x|\leq2),\qquad c(x)=0\quad(|x|>2).
\label{eq84}
\end{equation}
This is an exact convolution of the switching functions. It is nonnegative, even, zero at $x=0,\pm2$, and smooth enough at the support endpoints for the principal value manipulations below.

The resulting compact kernel gives the commutator contribution directly.
Equations~\eqref{eq28} and~\eqref{eq84} give
\begin{equation}
\cQ(Ty)=-\frac{c(y-\ell)+c(y+\ell)}{8\pi\ell}.
\end{equation}
For $\ell>2$ and $y\geq0$, the second term vanishes. Therefore
\begin{equation}
|\cQ(Ty)|=\begin{cases}
\displaystyle\frac{\sin^5(\pi|y-\ell|/2)}{120\pi^2\ell},&|y-\ell|\leq2,\\
0,&|y-\ell|>2.
\end{cases}
\label{eq86}
\end{equation}
The commutator response has two maxima, at $y=\ell-1$ and $y=\ell+1$, both equal to $1/(120\pi^2\ell)$. Between them it vanishes at $y=\ell$, because the oscillatory detector response gives $c(0)=0$. The interaction regions still meet the light cone at this point. These maxima refer to $|\cQ|$. The total payoff $g(s)$ also contains the delay-dependent Hadamard term and local excitation noise.

To determine whether the resulting coherence gives a positive payoff, we must also calculate the local excitation noise. At a single spatial point, the vacuum mode expansion gives
\begin{equation}
W(t,0)=\frac1{4\pi^2}\int_0^\infty k\ee^{-\ii k(t-\ii0)}\dd k.
\end{equation}
Substituting this distribution into Eq.~\eqref{eq8}, the time integrations factorize
\begin{align}
\cP&=\frac1{4\pi^2}\int_0^\infty k\dd k
\left[\int\chi(t)\ee^{-\ii(\Omega+k)t}\dd t\right]
\left[\int\chi(t')\ee^{\ii(\Omega+k)t'}\dd t'\right]\nonumber\\
&=\frac1{4\pi^2}\int_0^\infty k|\widetilde\chi(\Omega+k)|^2\dd k.
\label{eq88}
\end{align}
Substituting Eq.~\eqref{eq83} into Eq.~\eqref{eq88} gives
\begin{equation}
\cP(\Omega T)=\frac{\pi^2}{4}\int_0^\infty
\frac{k\sin^2[(\Omega+k)T]}{(\Omega+k)^2[\pi^2-(\Omega+k)^2T^2]^2}\dd k.
\label{eq89}
\end{equation}
It is finite. At infinity the integrand is $\ord(k^{-5})$, and the apparent finite frequency poles are removable.

At $\Omega T=3\pi/2$, the delay independent local noise is obtained directly from the convergent one-dimensional integral in Eq.~\eqref{eq89}. Standard high precision quadrature gives
\begin{equation}
\cP(3\pi/2)=0.0001984768304305\ldots.
\end{equation}
Below, $\cP$ denotes the local noise at this selected gap. This fixed noise floor is all that is needed below, so we use the convergent integral directly.

The nonlocal Hadamard contribution requires a separate principal value integral.
The real part is
\begin{equation}
\cH(Ty)=\frac1{4\pi^2}\PV\int_{-2}^{2}
\frac{c(x)}{\ell^2-(x-y)^2}\dd x.
\label{eq91}
\end{equation}
Evenness of $c$ does not imply that the entire integrand is even for nonzero $y$. Substituting $x\mapsto-x$ in the negative half, we have
\begin{equation}
\cH(Ty)=\frac1{4\pi^2}\PV\int_0^2c(x)
\left[\frac1{\ell^2-(x-y)^2}+\frac1{\ell^2-(x+y)^2}\right]\dd x.
\label{eq92}
\end{equation}
At $\Omega T=3\pi/2$,
\begin{equation}
c(x)=\frac{10\sin(\pi x/2)-5\sin(3\pi x/2)+\sin(5\pi x/2)}{240\pi}
\quad(0\leq x\leq2).
\end{equation}
For positive odd $n$, define
\begin{equation}
K_n(a)=\PV\int_0^2\sin\!\left(\frac{n\pi x}{2}\right)\left(\frac1{a-x}+\frac1{a+x}\right)\dd x.
\end{equation}
Using $[\ell^2-u^2]^{-1}=[(\ell-u)^{-1}+(\ell+u)^{-1}]/(2\ell)$ in Eq.~\eqref{eq92},
\begin{equation}
\begin{aligned}
\cH(Ty)=\frac1{1920\pi^3\ell}\Bigl\{&
10[K_1(\ell+y)+K_1(\ell-y)]-5[K_3(\ell+y)+K_3(\ell-y)]\\
&+[K_5(\ell+y)+K_5(\ell-y)]\Bigr\}.
\end{aligned}
\end{equation}

The principal value kernel admits a closed endpoint form that also exposes the spacelike ordering.
We use the real sine and cosine integrals
\begin{equation}
\Si(z)=\int_0^z\frac{\sin t}{t}\dd t,
\qquad
\Ci(z)=\gamma+\ln z+\int_0^z\frac{\cos t-1}{t}\dd t\quad(z>0).
\end{equation}
Here $\gamma$ is the Euler-Mascheroni constant.
For $x\ne a$, integrating $\sin(n\pi x/2)/(a-x)$ with respect to $x$ gives
\begin{equation}
-\sin\!\left(\frac{n\pi a}{2}\right)\Ci\!\left(\frac{n\pi|x-a|}{2}\right)
-\cos\!\left(\frac{n\pi a}{2}\right)\Si\!\left(\frac{n\pi(x-a)}{2}\right),
\end{equation}
Differentiating this expression with respect to $x$ recovers the integrand, using $\dd\Ci(|z|)/\dd z=\cos z/z$ for nonzero real $z$. Similarly, for $x\ne-a$, integrating $\sin(n\pi x/2)/(a+x)$ gives
\begin{equation}
\cos\!\left(\frac{n\pi a}{2}\right)\Si\!\left(\frac{n\pi(x+a)}{2}\right)
-\sin\!\left(\frac{n\pi a}{2}\right)\Ci\!\left(\frac{n\pi|x+a|}{2}\right).
\end{equation}
The logarithmic contributions on the two sides of an interior pole cancel under symmetric exclusion. Evaluating the endpoints gives
\begin{align}
K_n(a)={}&\sin\!\left(\frac{n\pi a}{2}\right)
\left[2\Ci\!\left(\frac{n\pi|a|}{2}\right)-\Ci\!\left(\frac{n\pi|a-2|}{2}\right)-\Ci\!\left(\frac{n\pi|a+2|}{2}\right)\right]\nonumber\\
&+\cos\!\left(\frac{n\pi a}{2}\right)
\left[\Si\!\left(\frac{n\pi(a+2)}{2}\right)+\Si\!\left(\frac{n\pi(a-2)}{2}\right)-2\Si\!\left(\frac{n\pi a}{2}\right)\right].
\end{align}
The function is odd in $a$. The explicit continuous limits for odd $n$ are
\begin{equation}
K_n(0)=0,\qquad K_n(2)=2\Si(n\pi)-\Si(2n\pi),\qquad K_n(-2)=-K_n(2).
\end{equation}
Terms of the form $z\ln|z|$ tend to zero. These limits must be used rather than evaluating a formal product of zero and a divergent $\Ci$ in floating point arithmetic.

At $\tau=0$, the ordering of the strictly spacelike timing payoffs follows directly from the shape of $\cH$. For $|y|<\ell-2$, Eq.~\eqref{eq91} has no pole, and differentiating twice gives
\begin{equation}
\frac{\dd^2}{\dd y^2}\cH(Ty)
=\frac1{2\pi^2}\int_{-2}^2
\frac{c(x)[\ell^2+3(x-y)^2]}{[\ell^2-(x-y)^2]^3}\dd x>0.
\end{equation}
The denominator is positive, and the kernel $c(x)$ is nonnegative and not identically zero. Thus $\cH$ is positive and strictly convex in this interval. Since it is also even, its minimum occurs at zero delay, so
\begin{equation}
\cH(-\delta)=\cH(\delta)>\cH(0)\quad\text{for }0<\frac{\delta}{T}<\ell-2.
\end{equation}
Since $\cQ=0$, the reward is $g(s)=\Gamma\lambda^2[\cH(s)-\cP]$, giving $g_-=g_+>g_0$ at $\tau=0$. If $\cH(0)>\cP$, even the smallest reward $g_0$ is positive. The anti-coordination case in Sec.~\ref{sec4} then gives three positive-reward active equilibria and hence seven full equilibria. The next subsection verifies $\cH(0)>\cP$ for a specific separation.

\subsection{Equilibria at different delays}
Keeping $\Omega T=3\pi/2$ as above, we now fix the separation and timing menu by choosing
\begin{equation}
\ell=\frac{21}{10},\qquad \frac{\delta}{T}=\frac9{100}.
\end{equation}
The choice $L/T=2.1>2$ ensures that simultaneous switching intervals are strictly spacelike separated. Evaluating the response functions then gives $\cH(0)\simeq2.2493\times10^{-4}>\cP\simeq1.9848\times10^{-4}$, so the simultaneous history has a positive payoff. The menu spacing $\delta/T=0.09$ distinguishes the early and late histories while satisfying $0<\delta<L-2T$. This condition keeps every history strictly spacelike at $\tau=0$ and beyond the light cone shell at $\tau=2L$.

For these parameters, the largest time difference between the two interaction regions at $\tau=0$ is $2T+\delta=2.09T<L=2.1T$. At $\tau=2L$, the smallest difference is $(2\ell-2)T-\delta=2.11T>L$. Thus every history in the first setting is strictly spacelike, and every history in the last is strictly beyond the light cone shell. Both have an exact clearance of $0.01T$.

The five baseline delays are selected by the analytic causal kernel,
\begin{equation}
\frac\tau T\in\{0,\ell-1,\ell,\ell+1,2\ell\}
=\left\{0,\frac{11}{10},\frac{21}{10},\frac{31}{10},\frac{42}{10}\right\}.
\end{equation}
The setting $\tau=0$ supplies the strictly spacelike reference. The delays $\tau=L-T$ and $L+T$ sample the two maxima of $|\cQ|$ in Eq.~\eqref{eq86}, while $\tau=L$ selects its central carrier cancellation, $\cQ(\tau)=0$. Finally, $\tau=2L$ puts every timing history beyond the wavefront, where Huygens propagation makes $\cQ=0$. The five settings thus compare distinct causal features of one fixed protocol, rather than extrema of its total payoff. Table~\ref{tab3} gives the field contributions at the fifteen relative delays entering these five games. We evaluate the analytic formulas above. The last column reports $g(s)/(\Gamma\lambda^2)$, so the same table applies to any fixed positive payment scale and nonzero coupling.

\begingroup\small
\begin{table}[H]
\centering
\caption{The fifteen field histories, rounded to twelve decimal places from high-precision evaluation. }\label{tab3}
\begin{tabular}{c c c c}
\toprule $s/T$&$\cH$&$|\cQ|$&$g(s)/(\Gamma\lambda^2)$\\\midrule
$-0.09$&0.000226654368&0&0.000028177537\\
0&0.000224934488&0&0.000026457657\\
0.09&0.000226654368&0&0.000028177537\\\midrule
1.01&0.000205317716&0.000382408535&0.000235564242\\
1.10&0.000105059192&0.000402068189&0.000217090568\\
1.19&0.000005161134&0.000382408535&0.000183966532\\\midrule
2.01&0.000024347965&0.000000022329&-0.000174128855\\
2.10&0.000044094118&0&-0.000154382713\\
2.19&0.000063895204&0.000000022329&-0.000134581622\\\midrule
3.01&0.000089104091&0.000382408535&0.000194175467\\
3.10&-0.000009402560&0.000402068189&0.000203701285\\
3.19&-0.000108080184&0.000382408535&0.000198911662\\\midrule
4.11&-0.000093403179&0&-0.000105073652\\
4.20&-0.000084110035&0&-0.000114366796\\
4.29&-0.000076523526&0&-0.000121953304\\
\bottomrule
\end{tabular}
\end{table}
\endgroup

Using the timing rewards $g_0=g(\tau)$ and $g_\pm=g(\tau\pm\delta)$, sufficient strict inequalities are
\begin{equation}
\begin{aligned}
\tau/T=0:\quad &g_0>2.6\times10^{-5}\Gamma\lambda^2,\quad
g_- -g_0=g_+-g_0>1.7\times10^{-6}\Gamma\lambda^2;\\
\tau/T=1.1:\quad &\min(g_-,g_0,g_+)>1.8\times10^{-4}\Gamma\lambda^2,\\
&g_- -g_0>1.8\times10^{-5}\Gamma\lambda^2,\quad
g_0-g_+>3.3\times10^{-5}\Gamma\lambda^2;\\
\tau/T=2.1:\quad &\max(g_-,g_0,g_+)<-1.3\times10^{-4}\Gamma\lambda^2;\\
\tau/T=3.1:\quad &\min(g_-,g_0,g_+)>1.9\times10^{-4}\Gamma\lambda^2,\\
&g_0-\max(g_-,g_+)>4.7\times10^{-6}\Gamma\lambda^2;\\
\tau/T=4.2:\quad &\max(g_-,g_0,g_+)<-10^{-4}\Gamma\lambda^2.
\end{aligned}
\end{equation}
At $\tau=0$, the equality $g_-=g_+$ follows from exact evenness. The stricter orderings for the two all-negative matrices can also be read from Table~\ref{tab3}, though negativity alone already proves unique exit.

\begin{table}[H]
\centering
\caption{Five settings of the same protocol. Rewards $(g_-,g_0,g_+)$ are shown in units of $10^{-4}\Gamma\lambda^2$. None means that no active equilibrium has positive reward.}
\label{tab4}
\begin{tabular}{c c c c}
\toprule
$\tau/T$ & $(g_-,g_0,g_+)$ & Active type & $N_{\rm THP}$\\
\midrule
0 & $(0.2818,0.2646,0.2818)$ & Anti-coordination & 7\\
1.1 & $(2.3556,2.1709,1.8397)$ & Directional & 3\\
2.1 & $(-1.7413,-1.5438,-1.3458)$ & None & 1\\
3.1 & $(1.9418,2.0370,1.9891)$ & Coordination & 7\\
4.2 & $(-1.0507,-1.1437,-1.2195)$ & None & 1\\
\bottomrule
\end{tabular}
\end{table}

The five field evaluations summarized in Table~\ref{tab4} now determine the equilibrium sequence directly.
At $\tau/T=0$, the active game anti-coordinates. Its pure equilibria are $(0,1)$ and $(1,0)$. Symmetry gives $x_*=y_*=1/2$ and $u_*=(g_0+g_+)/2>0$. All three active equilibria have positive rewards, so the full count is seven. This first setting uses a strictly spacelike field resource.

At $\tau/T=1.1$, the ordering is $g_->g_0>g_+>0$. Alice strictly prefers late and Bob strictly prefers early, so only $(1,0)$ survives in the active game. Its reward $g_->0$ gives a certain-entry equilibrium and a mixed-entry equilibrium, which together with mutual exit give three.

At $\tau/T=2.1$, every entry of the active matrix is negative. Any mixed active reward, being a convex combination of matrix entries, is negative as well. Exit strictly dominates participation, so the count is one. The central $\cQ$ vanishes, but the neighboring histories have a small nonzero commutator contribution and the regions remain causally connected. This is not a Huygens post-contact example.

At $\tau/T=3.1$, $g_0>\max(g_-,g_+)>0$. There are two coordinated pure equilibria and one interior mixed equilibrium, and all have positive reward because every active matrix entry is positive. Each gives a certain-entry equilibrium and a mixed-entry equilibrium, yielding seven when mutual exit is included. The mixed timing probabilities need not be equal, since $g_-\ne g_+$.

At $\tau/T=4.2$, all sampled histories are strictly timelike and beyond the commutator shell. Thus $\cQ=0$ exactly. The remaining $|\cH|$ is smaller than $\cP$ at each sample, making all rewards negative and leaving one perfect equilibrium. Huygens support proves the absence of the commutator term, and the inequality involving $\cH$ is separately needed to prove the exit result.

The different equilibria at $\tau=L-T$ and $\tau=L+T$ come from the Hadamard contribution. For the chosen switching function, the three sampled values of $|\cQ|$ are identical at the two delays, and $\cP$ is unchanged. In contrast, $\cH$ involves the principal value integral over all pairs of interaction times, as shown in Eq.~\eqref{eq91}. Its kernel has no reflection symmetry about $s=L$, so $\cH$ need not take equal values at $L-T$ and $L+T$. The different Hadamard values give $g_->g_0>g_+$ at the first lobe and $g_0>\max(g_-,g_+)$ at the second.

\section{Discussion and conclusions}\label{sec6}

In this work, we show that entanglement obtained by localized observers from a quantum field can also influence their behavior as strategic agents. Entanglement harvesting has shown that initially uncorrelated detectors may become entangled through their coupling to a quantum field even when they are spacelike separated. We take a further operational step by embedding the detector state in a fixed quantum-input participation game. Within this protocol, the harvested detector state becomes the input to the observers' strategic interaction. The observers choose when to couple their detectors to the field and whether to participate at all. A fixed quantum input verification rule then assigns a payoff to the shared state they prepare. This turns the usual calculation of harvested entanglement into a question about which choices the two observers can maintain against one-sided deviations.

For this finite game, we obtain a complete analytic count. When $g_0\ne g_\pm$ and every active equilibrium has nonzero payoff, Section~\ref{sec4} gives
\[
N_{\rm THP}=1+2n_+\in\{1,3,5,7\},
\]
where $n_+$ counts the active timing equilibria with positive payoff. Each gives two full equilibria, with certain entry or with entry mixed with exit. Mutual exit supplies the remaining one.

The quantum field enters this count through the timing payoffs. For the detector state considered in this work, their leading value is proportional to $\sqrt{\cH^2+\cQ^2}-\cP$. The Hadamard and Pauli-Jordan terms contribute to the pair coherence, while $\cP$ is the local excitation noise. The coherence must exceed this noise before an active equilibrium can give a positive payoff. The field correlations thus determine both which timing choices are best responses and whether their reward can sustain participation. We give an explicit example of this general result by a free massless scalar field calculation. With the same switching function, gap, separation, and payment rule, the five delays in Table~\ref{tab4} give $7\to3\to1\to7\to1$. They sample strictly spacelike interactions, the first response lobe, the central cancellation, the second lobe, and the post-wavefront region. The counting theorem applies throughout the nondegenerate payoff regions. These five settings illustrate some of the cases reached by varying one delay.

At \(\tau=0\), all timing histories are strictly spacelike, so \(\mathcal Q=0\). The Hadamard contribution exceeds the local noise, and its even, convex profile favors opposite timing choices. At \(\tau=L-T\) and \(\tau=L+T\), the commutator gives the same three sampled magnitudes, but the Hadamard contributions differ. The resulting timing equilibria are directional at the first lobe and coordinated at the second. Thus, the seven full equilibria at \(\tau=0\) and \(\tau=L+T\) have different timing structures. The two settings with only mutual exit also have different physical origins. At \(\tau=L\), cancellation in the detector response gives \(\mathcal Q(L)=0\), while the neighboring histories retain small commutator contributions. At \(\tau=2L\), every history lies beyond the commutator support, so \(\mathcal Q=0\) throughout the timing menu by the massless Huygens property. For the chosen parameters, the coherence remains below the local noise in every history of both settings, leaving mutual exit as the only equilibrium.

Some readers may wonder what role entanglement actually plays here, given that the players' payoffs are assigned solely on the basis of the payoff table and Bell measurements. This point can be seen from the case of two players who share no entanglement. For two players without entanglement, the state between them can be written as a separable state. Under the same trusted questions, payment coefficients, and communication restrictions, every separable replacement gives a nonpositive payoff, even with arbitrary local binary measurements. A positive active payoff therefore certifies the use of detector entanglement in this particular game. Classical game theory could reproduce the same numerical matrices by assigning new payments, but that would change the verification problem. This quantum input verification does not require the harvested state to violate a Bell inequality. With trusted question preparation and the stated communication restrictions, its separable bound remains valid independently of the implementation of the local readouts~\cite{Buscemi2012,Branciard2013}.

We have given a simple, fully specified model where entanglement harvested from a quantum field becomes a resource for a quantum game. The observers choose to exit or to an early or late interaction before state preparation. If both enter, their detectors interact with the field at the chosen times to prepare the shared state used in the game. The relation between field correlations, verified payoffs, and Nash equilibria then determines which choices are mutually stable. We compute the detector state to leading perturbative order and solve the resulting finite timing and participation game exactly for these payoffs. The question states, payment rule, phase calibration, and Bell readouts are fixed throughout this game. Appendix~\ref{secA} also shows that the Bell readouts are mutual best responses to leading order on a positive payoff branch, with the detector state fixed. This is one possible construction of a quantum game from harvested entanglement. Alternative verification rules or strategy spaces can lead to different games and different equilibrium structures.

It is natural to consider the following extensions. For example, a continuous timing menu would allow us to determine whether the coordination and anti-coordination patterns persist as isolated equilibria or become continuous branches. Optimizing both readouts would reveal what other measurement equilibria are possible and how they interact with the timing choices. A massive field provides a different physical extension. Its commutator has a timelike tail, so the causal contribution can persist after the wavefront has passed. The late time equilibrium would then depend on this tail as well as on the Hadamard correlations and local noise. Correlation harvesting has also been proposed for quantum superpositions of spacetime backgrounds~\cite{Tang2026}, where the same verification rule could be used to study how the prepared resource affects participation. We can also consider curved spacetime. On a globally hyperbolic background, Hadamard states share the local singularity structure of the two-point function, while the causal propagator is fixed by the geometry and the field equation~\cite{BrunettiFredenhagenVerch2003,HollandsWald2015}. We could then compare changes due to the field state with those due to propagation in the background geometry. This would let the timing game probe how curvature, horizons, or cosmological expansion affect the resource carried by the detectors. Related operational approaches also appear in holography, where traversable wormholes have been analyzed as quantum channels and entanglement witnesses~\cite{Bao2018}. We will continue to investigate these questions in future work.

\appendix
\numberwithin{equation}{section}
\renewcommand{\theequation}{\thesection\arabic{equation}}

\section{Best response to a Bell readout}\label{secA}

So far the Bell readout has been prescribed as part of the verification rule. We now ask whether a player could improve the payoff by replacing only their own measurement. We keep the detector state fixed and let the other player retain the Bell readout. This isolates the measurement question from the timing and entry choices. The calculation will show that, on a positive leading-order payoff branch, the Bell-Bell pair is already a Nash equilibrium of this readout game.

Keep Bob's answer-$1$ effect fixed at $\Phi^+_{BQ_B}$. Alice may choose any answer-$1$ effect $A_1$ on $Q_A\otimes A$, with $0\leq A_1\leq\id_{Q_AA}$. The detector state, denoted by $\rho=\rho_{AB}$, is a general X state
\begin{equation}
\rho=\begin{pmatrix}a&0&0&m^*\\0&b&l^*&0\\0&l&c&0\\m&0&0&d\end{pmatrix}.
\end{equation}
Here $a,b,c,d$ are real diagonal entries, while $m,l$ are the two generally complex coherences in the detector basis $(\ket{gg},\ket{ge},\ket{eg},\ket{ee})$.

The only change from Eq.~\eqref{eq46} is the replacement
$\Phi^+_{Q_AA}\mapsto A_1$ on Alice's side. For questions $q,r$, the probability of the rewarded outcome becomes
\begin{equation}
p_{A_1,\Phi}(1,1|q,r)=
\Tr\!\left[(A_1^{Q_AA}\otimes\Phi^+_{BQ_B})
(\tau_q^{Q_A}\otimes\rho_{AB}\otimes\tau_r^{Q_B})\right].
\label{eqA2}
\end{equation}

Define the unnormalized conditional operator on $A$
\begin{align}
R_A^{(r)}
&=\Tr_{BQ_B}\!\left[(\id_A\otimes\Phi^+_{BQ_B})
(\rho_{AB}\otimes\tau_r^{Q_B})
(\id_A\otimes\Phi^+_{BQ_B})\right]\nonumber\\
&=\frac12\Tr_B\!\left[\rho_{AB}(\id_A\otimes\tau_r^T)\right].
\label{eqA3}
\end{align}
The second line follows from Eq.~\eqref{eq45}. Using the Bloch components of $\tau_r$ already defined in Sec.~\ref{sec3}, direct evaluation gives
\begin{equation}
R_A^{(r)}=\frac14
\begin{pmatrix}
a(1+r_{rz})+b(1-r_{rz})&m^*(r_{rx}-\ii r_{ry})+l^*(r_{rx}+\ii r_{ry})\\
l(r_{rx}-\ii r_{ry})+m(r_{rx}+\ii r_{ry})&c(1+r_{rz})+d(1-r_{rz})
\end{pmatrix}.
\end{equation}
The off diagonal entries show explicitly why the full two detector state is required. Equation~\eqref{eqA2} now reads
\begin{equation}
p_{A_1,\Phi}(1,1|q,r)
=\Tr_{Q_AA}\!\left[A_1(\tau_q\otimes R_A^{(r)})\right].
\end{equation}

Both the Born probability and the mean payoff are linear in $A_1$. Hence all fixed ingredients can be collected into one effective Hermitian operator,
\begin{align}
G_A&=-4\Gamma\sum_{q,r}\beta_{qr}p_{A_1,\Phi}(1,1|q,r)
=-4\Gamma\Tr_{Q_AA}(A_1K_A),\nonumber\\
K_A&:=\sum_{q,r}\beta_{qr}\tau_q\otimes R_A^{(r)},
\label{eqA6}
\end{align}

We could obtain $K_A$ by inserting the sixteen coefficients $\beta_{qr}$ directly into Eq.~\eqref{eqA6}. It is useful to carry out the sums in stages, since the tetrahedral identities make the cancellations explicit.

Using $\tau_q=(\id+\bm r_q\cdot\bm\sigma)/2$, these two geometric identities give
\begin{equation}
\sum_q\tau_q=2\id,
\qquad
\sum_q r_{qj}\tau_q=\frac23\sigma_j,
\qquad j=x,y,z.
\end{equation}
Taking the transpose gives the corresponding identities for Bob's contracted question states,
\begin{equation}
\sum_r\tau_r^T=2\id,
\qquad
\sum_r r_{rj}\tau_r^T=\frac23\sigma_j^T.
\end{equation}
Insert Eq.~\eqref{eqA3} into the definition of $K_A$. Before using the explicit $\beta_{qr}$, this gives
\begin{equation}
K_A=\frac12\sum_{q,r}\beta_{qr}\tau_q\otimes
\Tr_B[\rho_{AB}(\id_A\otimes\tau_r^T)].
\end{equation}
Equation~\eqref{eq42} separates this double sum into four products of one-index sums:
\begin{equation}
\begin{aligned}
K_A=\frac1{32}\Bigg\{&
\left(\sum_q\tau_q\right)\otimes
\Tr_B\!\left[\rho_{AB}\left(\id_A\otimes\sum_r\tau_r^T\right)\right]\\
&-9\left(\sum_qr_{qx}\tau_q\right)\otimes
\Tr_B\!\left[\rho_{AB}\left(\id_A\otimes\sum_rr_{rx}\tau_r^T\right)\right]\\
&+9\left(\sum_qr_{qy}\tau_q\right)\otimes
\Tr_B\!\left[\rho_{AB}\left(\id_A\otimes\sum_rr_{ry}\tau_r^T\right)\right]\\
&-9\left(\sum_qr_{qz}\tau_q\right)\otimes
\Tr_B\!\left[\rho_{AB}\left(\id_A\otimes\sum_rr_{rz}\tau_r^T\right)\right]
\Bigg\}.
\end{aligned}
\end{equation}
Applying these identities gives
\begin{equation}
\begin{aligned}
K_A=\frac18\Bigl\{&\id\otimes\rho_A
-\sigma_x\otimes\Tr_B[\rho_{AB}(\id_A\otimes\sigma_x^T)]\\
&+\sigma_y\otimes\Tr_B[\rho_{AB}(\id_A\otimes\sigma_y^T)]
-\sigma_z\otimes\Tr_B[\rho_{AB}(\id_A\otimes\sigma_z^T)]\Bigr\},
\end{aligned}
\end{equation}
where the required partial traces are
\begin{align}
\rho_A&=\begin{pmatrix}a+b&0\\0&c+d\end{pmatrix},
\nonumber\\
\Tr_B[\rho_{AB}(\id_A\otimes\sigma_x^T)]&=\begin{pmatrix}0&m^*+l^*\\m+l&0\end{pmatrix},\nonumber\\
\Tr_B[\rho_{AB}(\id_A\otimes\sigma_y^T)]&=\begin{pmatrix}0&\ii(l^*-m^*)\\\ii(m-l)&0\end{pmatrix},\nonumber\\
\Tr_B[\rho_{AB}(\id_A\otimes\sigma_z^T)]&=\begin{pmatrix}a-b&0\\0&c-d\end{pmatrix}.
\end{align}
Substitution yields the effective operator
\begin{equation}
K_A=\frac14\begin{pmatrix}
b&0&0&-m^*\\
0&d&-l&0\\
0&-l^*&a&0\\
-m&0&0&c
\end{pmatrix},
\label{eqA13}
\end{equation}
in the ordered basis $(\ket{g}_{Q_A}\ket{g}_A,\ket{g}_{Q_A}\ket{e}_A,\ket{e}_{Q_A}\ket{g}_A,\ket{e}_{Q_A}\ket{e}_A)$. This space is distinct from the detector space $A\otimes B$, although both use four basis vectors.

Because $G_A=-4\Gamma\Tr(A_1K_A)$, maximizing the reward is equivalent to minimizing $\Tr(A_1K_A)$. In the eigenbasis of $K_A$, each diagonal element of $A_1$ lies between zero and one. The minimum is therefore obtained by choosing one on the negative eigenvalue directions and zero on the positive ones. It equals the sum of the negative eigenvalues of $K_A$ and is attained by projecting onto its negative spectral subspace.

From this point onward, $\rho$ and $\rho_{AB}$ denote the fixed calibrated resource $\widetilde\rho_{AB}$ of the main text. For identical detectors,
\begin{equation}
\begin{alignedat}{2}
b&=c=\lambda^2\cP+\ord(\lambda^4), &\qquad
m&=\lambda^2|Z|+\ord(\lambda^4),\\
a&=1-2\lambda^2\cP+\ord(\lambda^4), &\qquad
l&=\ord(\lambda^2),\quad
d=\ord(\lambda^4).
\end{alignedat}
\end{equation}
Substituting into Eq.~\eqref{eqA13}, the four eigenvalues of $K_A$ become
\begin{equation}
\frac14+\ord(\lambda^2),\qquad \ord(\lambda^4),\qquad
\frac{\lambda^2}{4}(\cP\pm|Z|)+\ord(\lambda^4).
\end{equation}
The only eigenvalue that can be negative at order $\lambda^2$ is proportional to $\cP-|Z|$. When $|Z|>\cP$, its eigenvector is $\ket{\Phi^+}_{Q_AA}$. Thus Alice's optimal answer-$1$ effect is the Bell projector to this order. The other block of $K_A$ can affect the optimum only from order $\lambda^4$ onward.

With this choice, the minimum is attained by the original Bell readout,
\begin{equation}
\Tr(\Phi^+_{Q_AA}K_A)
=\frac{b+c-m-m^*}{8}
=\frac14\Tr(W_0\rho_{AB}).
\end{equation}
Consequently, on the positive payoff branch, the maximal one-sided reward coincides with the payoff in Eq.~\eqref{eq51} to leading order,
\begin{equation}
\max_{A_1}G_A=-\Gamma\Tr(W_0\rho_{AB})+\ord(\lambda^4)
=\Gamma\lambda^2(|Z|-\cP)+\ord(\lambda^4).
\end{equation}

The same argument applies with Alice and Bob exchanged. Each Bell measurement is therefore a leading-order best response to the other on the positive margin branch. This proves the stated Bell-Bell equilibrium for a fixed detector state.

\section{Trembling hand perfection}\label{secB}
This appendix proves perfection of the Nash equilibria counted in Sec.~\ref{sec4}. For a finite normal-form game, an equilibrium is trembling hand perfect if there is a sequence of completely mixed profiles converging to it, against which each player's original equilibrium strategy remains a best response~\cite{Selten1975}. Here \emph{completely mixed} means that exit, early entry, and late entry all have positive probability. The perturbed profiles themselves need not be Nash equilibria.

For each case, we first state the original equilibrium, then specify the opponent's perturbed probabilities, and finally compare the available payoffs. A pure equilibrium action must remain preferable to the alternatives. For an original mixed strategy, its supported actions must remain tied for the highest payoff. We use $\varepsilon\to0^+$ separately in each construction. The perturbation may represent accidental entry, accidental exit, or an unintended timing choice.

Unless stated otherwise, probability triples are ordered as $(E,0,1)$, denoting exit, early entry, and late entry. We write $g_{00}=g_{11}=g_0$, $g_{01}=g_+$, and $g_{10}=g_-$. Alice's payoff from a pure action against Bob's specified strategy is denoted by $U_A$, with $U_A(E)=0$; Bob's is defined analogously.

\emph{Mutual exit.} The original equilibrium is
\begin{equation}
\sigma_A=\sigma_B=(1,0,0).
\end{equation}
Let Bob instead choose
\begin{equation}
\sigma_B^{(\varepsilon)}=(1-2\varepsilon,\varepsilon,\varepsilon),
\qquad 0<\varepsilon<\frac12.
\end{equation}
He now enters accidentally with probability $\varepsilon$ at each timing. Against this strategy, Alice's payoff from either entrance action is
\begin{equation}
U_A(i)=-(1-2\varepsilon)
+\varepsilon(g_{i0}+g_{i1})\longrightarrow-1<0=U_A(E),
\qquad i=0,1.
\end{equation}
Thus keeping her original choice of exit is strictly better than entering when $\varepsilon$ is sufficiently small. Giving Alice the same perturbed probabilities, the same argument makes Bob's original choice optimal.

\emph{Certain entry with pure timing choices.} Let $(i,j)$ be a strict active equilibrium with common reward $u=g_{ij}>0$. Write $\bar i=1-i$ and $\bar j=1-j$ for the alternative timings. Strictness means $g_{ij}>g_{\bar i j}$ and $g_{ij}>g_{i\bar j}$. In the action orders $(E,i,\bar i)$ for Alice and $(E,j,\bar j)$ for Bob, the original strategies are
\begin{equation}
\sigma_A=\sigma_B=(0,1,0).
\end{equation}
Let Bob's perturbed strategy, in his stated action order, be
\begin{equation}
\sigma_B^{(\varepsilon)}=(\varepsilon,1-2\varepsilon,\varepsilon),
\qquad 0<\varepsilon<\frac12.
\end{equation}
He has a small probability of exiting or choosing the unintended timing. Alice's original timing then gives
\begin{equation}
U_A(i)=-\varepsilon+(1-2\varepsilon)g_{ij}
+\varepsilon g_{i\bar j}\longrightarrow u>0=U_A(E).
\end{equation}
Her gain over switching to the alternative timing is
\begin{equation}
\begin{aligned}
U_A(i)-U_A(\bar i)
&=(1-2\varepsilon)(g_{ij}-g_{\bar i j})
+\varepsilon(g_{i\bar j}-g_{\bar i\bar j})\\
&\longrightarrow g_{ij}-g_{\bar i j}>0.
\end{aligned}
\end{equation}
Both inequalities remain strict for sufficiently small $\varepsilon$, so Alice should retain her original timing rather than switch or exit. Perturb Alice in the same way, in her own action order. The condition $g_{ij}>g_{i\bar j}$ gives the corresponding conclusion for Bob, proving perfection of the original pure equilibrium.

\emph{Certain entry with mixed timing choices.} Let $(x_*,y_*)$ be an interior active equilibrium with common reward $u>0$. Both players enter with certainty but randomize their timings, so the original strategies are
\begin{equation}
\begin{aligned}
\sigma_A&=(0,x_*,1-x_*),\\
\sigma_B&=(0,y_*,1-y_*),
\end{aligned}
\qquad 0<x_*,y_*<1.
\end{equation}
Let Bob exit accidentally with probability $\varepsilon$, retaining his original timing probabilities conditional on entry:
\begin{equation}
\sigma_B^{(\varepsilon)}
=(\varepsilon,(1-\varepsilon)y_*,
(1-\varepsilon)(1-y_*)).
\end{equation}
Because $A_0(y_*)=A_1(y_*)=u$, Alice's two timings still give the same payoff,
\begin{equation}
U_A(0)=U_A(1)=(1-\varepsilon)u-\varepsilon>0=U_A(E)
\quad\text{if}\quad
0<\varepsilon<\frac{u}{1+u}.
\end{equation}
Consequently, Alice's original mixture remains optimal. Changing the early/late probabilities cannot improve her payoff, while assigning positive probability to exit would lower it. Give Alice the perturbed strategy $(\varepsilon,(1-\varepsilon)x_*,(1-\varepsilon)(1-x_*))$. The same reasoning applies to Bob, since $B_0(x_*)=B_1(x_*)=u$. All actions then have positive probability, and both original mixed strategies remain best responses as $\varepsilon\to0^+$.

\emph{Mixed entry with pure timing choices.} Fix a strict active equilibrium $(i,j)$ with $g_{ij}>0$. Here $u=g_{ij}$ is the common payoff when both players enter and use their intended timings. In the action orders $(E,i,\bar i)$ and $(E,j,\bar j)$, the original strategies are
\begin{equation}
\sigma_A=\sigma_B=
\left(\frac{u}{1+u},\frac{1}{1+u},0\right).
\end{equation}
Each player enters with probability $1/(1+u)$ and, upon entry, uses the intended timing with certainty. Against this strategy, Alice's payoff from entry is $u/(1+u)-u/(1+u)=0$, equal to her exit payoff. Bob has the same indifference, so their expected payoffs in the participation equilibrium are zero even though $u>0$.

We now perturb Bob by giving his alternative timing conditional probability $\varepsilon>0$ upon entry. If Alice retains timing $i$, her reward conditional on joint entry becomes $(1-\varepsilon)u+\varepsilon g_{i\bar j}$. This generally differs from $u$, although it converges to $u$ as $\varepsilon\to0^+$. To preserve Alice's indifference between entry and exit, Bob's entry probability must therefore be adjusted. Requiring $\beta_{\varepsilon}[(1-\varepsilon)u+\varepsilon g_{i\bar j}]-(1-\beta_{\varepsilon})=0$ determines it. Applying the same condition to Bob determines Alice's entry probability:
\begin{equation}
\begin{aligned}
\alpha_{\varepsilon}&=\frac{1}{1+(1-\varepsilon)u+\varepsilon g_{\bar i j}},\\
\beta_{\varepsilon}&=\frac{1}{1+(1-\varepsilon)u+\varepsilon g_{i\bar j}},
\end{aligned}
\end{equation}
where $\alpha_{\varepsilon}$ and $\beta_{\varepsilon}$ are Alice's and Bob's entry probabilities. The perturbed strategies are
\begin{equation}
\begin{aligned}
\sigma_A^{(\varepsilon)}&=\bigl(1-\alpha_{\varepsilon},\alpha_{\varepsilon}(1-\varepsilon),\alpha_{\varepsilon}\varepsilon\bigr),\\
\sigma_B^{(\varepsilon)}&=\bigl(1-\beta_{\varepsilon},\beta_{\varepsilon}(1-\varepsilon),\beta_{\varepsilon}\varepsilon\bigr).
\end{aligned}
\end{equation}
For sufficiently small $\varepsilon$, all probabilities are positive because both conditional rewards approach $u>0$. Against Bob's perturbed strategy, Alice's originally supported actions remain tied:
\begin{equation}
U_A(i)=\beta_{\varepsilon}\bigl[(1-\varepsilon)u+\varepsilon g_{i\bar j}\bigr]
-(1-\beta_{\varepsilon})=0=U_A(E).
\end{equation}
Her unused timing is strictly worse, since
\begin{equation}
\begin{aligned}
U_A(i)-U_A(\bar i)
&=\beta_{\varepsilon}\bigl[(1-\varepsilon)(u-g_{\bar i j})
+\varepsilon(g_{i\bar j}-g_{\bar i\bar j})\bigr]\\
&\longrightarrow \frac{u-g_{\bar i j}}{1+u}>0.
\end{aligned}
\end{equation}
This follows from $u=g_{ij}>g_{\bar i j}$ at the original equilibrium. Alice's original mixture therefore remains a best response, and the same reasoning applies to Bob. 

\emph{Mixed entry with mixed timing choices.} Let $(x_*,y_*)$ be an interior active equilibrium with reward $u>0$. The original participation strategies are
\begin{equation}
\begin{aligned}
\sigma_A&=\left(\frac{u}{1+u},\frac{x_*}{1+u},\frac{1-x_*}{1+u}\right),\\
\sigma_B&=\left(\frac{u}{1+u},\frac{y_*}{1+u},\frac{1-y_*}{1+u}\right).
\end{aligned}
\end{equation}
All three actions already have positive probability for each player, so no additional perturbation is needed. Against Bob's strategy, Alice obtains
\begin{equation}
\begin{aligned}
U_A(E)&=0,&
U_A(0)&=\frac{A_0(y_*)-u}{1+u}=0,&
U_A(1)&=\frac{A_1(y_*)-u}{1+u}=0,
\end{aligned}
\end{equation}
because $A_0(y_*)=A_1(y_*)=u$. Her original mixture is therefore a best response, and changing any probabilities cannot improve her payoff. Bob satisfies the analogous equalities.

Together with the completeness argument in Sec.~\ref{sec4}, these constructions establish Eq.~\eqref{eq75}.

\section*{Acknowledgements}
Hao Xu thanks the National Natural Science Foundation of China (No.~12205250) for funding support.

\end{document}